# Learning the Imaging Landmarks: Unsupervised Key point Detection in Lung Ultrasound Videos

Arpan Tripathi, Mahesh Raveendranatha Panicker, Abhilash R Hareendranathan, Yale Tung Chen, Jacob L Jaremko, Kiran Vishnu Narayan and Kesavadas C

*Abstract—* Lung ultrasound (LUS) is an increasingly popular diagnostic imaging modality for continuous and periodic monitoring of lung infection, given its advantages of non-invasiveness, non-ionizing nature, portability and easy disinfection. The major landmarks assessed by clinicians for triaging using LUS are pleura, A and B lines. There have been many efforts for the automatic detection of these landmarks. However, restricting to a few pre-defined landmarks may not reveal the actual imaging biomarkers particularly in case of new pathologies like COVID-19. Rather, the identification of key landmarks should be driven by data given the availability of a plethora of neural network algorithms. This work is a first of its kind attempt towards unsupervised detection of the key LUS landmarks in LUS videos of COVID-19 subjects during various stages of infection. We adapted the relatively newer approach of transporter neural networks to automatically mark and track pleura, A and B lines based on their periodic motion and relatively stable appearance in the videos. Initial results on unsupervised pleura detection show an accuracy of 91.8% employing 1081 LUS video frames.

*Clinical Relevance—* This work establishes a novel way of unsupervised learning of imaging biomarkers which avoids the need for time-consuming image labeling, potentially allowing AI assessment of lung ultrasound to be broadly applied in the time of the COVID-19 pandemic.

## I. Introduction

Lung ultrasound (LUS) is an imaging modality highly suited for periodic repetitive monitoring of various pulmonary diseases. With the recent COVID-19 outbreak, automatic classification and scoring of lung ultrasound images to determine the severity of lung anomalies has gained momentum [1, 2]. The key landmarks in LUS such as pleura line (brightest and first horizontal line in Figs 3-6), A lines (horizontal lines as in Fig. 3) and B lines (vertical lines as in Fig. 5) and their alterations are indicative of lung involvement which can be used to screen high risk individuals who require hospitalization. There have been many attempts towards the usage of neural networks for automatic segmentation of these important landmarks and for tracking the prognosis of COVID-19 infection in subjects [3 -15]. These approaches rely on hand-crafted features which could be affected by noise and the presence of new pathologies (like COVID 19). Hence data driven approaches based on deep learning have also been tried for automatic detection of B lines as in [9, 10]. Instead of detection of pleura and A, B lines, there have been many approaches for classification of severity of lung infection as in [11 - 15]. However, all the approaches in [3 -15] assume predefined landmarks and are either fully supervised or semi-supervised approaches, which require significant amounts of data. This also requires significant effort and time from clinicians for annotations. Even in cases where large datasets are available there could be significant variability in annotations between readers leading to uncertainty in ground truth for supervised approaches.

Towards this, a novel unsupervised learning framework is proposed in this paper, which leverages the transporter architecture in [23] in order to learn key point representations for tracking prominent LUS features based on their periodic motion and relatively stable appearance in the video sequences.

## II. Proposed Methodology

In this section, the proposed approach is explained in detail with a brief review of the related work in computer vision (CV) and video processing community.

### A. Related Work

Unsupervised Learning has gained prominence in CV due to the potential to avoid the costly and time-consuming image labeling required in supervised learning. This has become possible with the utilization of innate properties of data, which include feature learning in images via reconstruction tasks [16] as well as the utilization of spatial dependency [17] and/or temporal dependency [18] between video frames to learn useful visual representations. Other approaches include learning suitable key points for tracking foreground objects by reconstruction of one video frame conditioned on another provided suitable difference in foreground object's pose in accordance to the quality of required key points [19, 20] and/or utilization of homographic warping on individual video frames [21]. Adaption of above methods for extraction of relevant key points in a given visual data population into LUS data through the suppression of irrelevant features [22] forms the basis of this work.

This work was supported by Department of Science and Technology - Science and Engineering Research Board (DSTSERB (CVD/2020/000221)) CRG COVID19 funding.

Arpan Tripathi and Mahesh Raveendranatha Panicker (email: mahesh@iitpkd.ac.in) are with Indian Institute of Technology Palakkad, Abhilash R Hareendranathan and Jacob L Jaremko are with University of Alberta Canada, Yale Tung Chen is with La Paz Hospital Spain, Kiran Vishnu Narayan is with Government Medical College Thiruvananthapuram, Kesavadas C is with Sree Chitra Tirunal Institute of Medical Sciences and Technology, Thiruvananthapuram

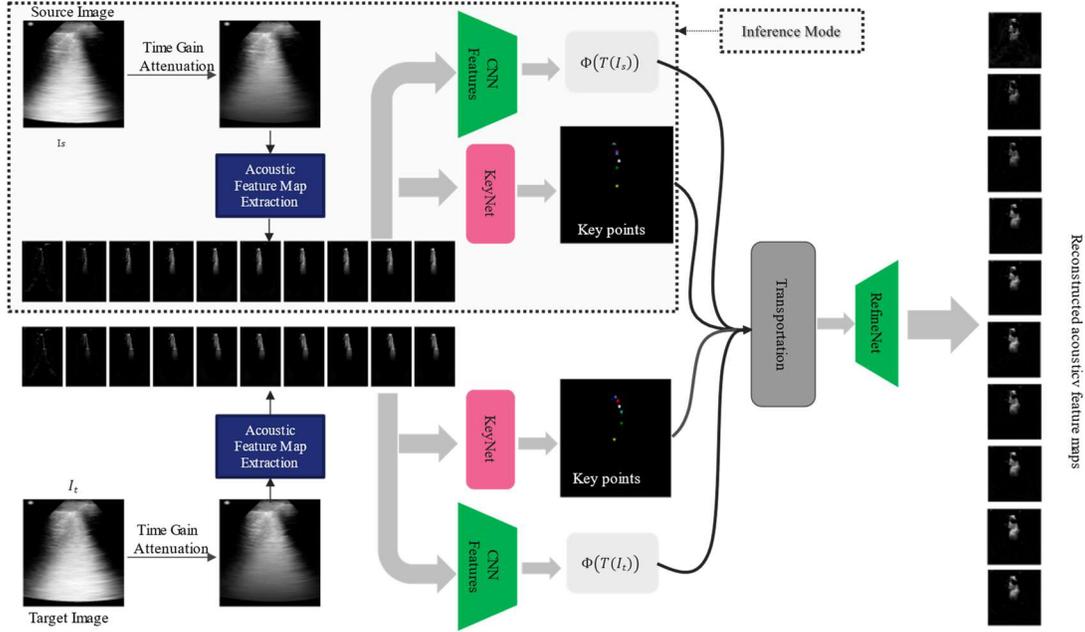

Fig. 1 Proposed key point detection network, based on [20]

## B. Proposed Architecture

In the proposed approach, the transporter architecture [20] is employed in order to learn key point representations for marking and tracking prominent LUS landmarks like Pleura, A and B lines based on their periodic motion and relatively stable appearance in the video sequences. The proposed architecture as derived from [20] is shown in Fig. 1. The major components of the proposed architecture are as described below.

**Time Gain Attenuation (TGA):** Each LUS frame in grayscale is resized to 256x256 resolution and pixelwise multiplied with a depth dependent decay mask as in $e^{-a \times d}$, where a is the attenuation factor and d is the depth. This is done to suppress the bright B patches which would otherwise cause hindrance in detection of pleural lines. Also, this in a way nullifies any time gain compensation applied on the acquired data. On the other hand, this could be controlled or removed in the case of B line identification.

**Acoustic Feature Fusion convolutional neural network (FF-CNN):** The FF-CNN is employed to learn the visual representations of the LUS sequences in a spatially smaller space but with more channels. The FF-CNN is pre-trained in an auto-encoder way to ensure it is extracting LUS relevant features. The steps in the acoustic feature fusion generation are as shown in Fig. 2. In Fig. 2, I is the input image, IBS(x,y)=$\sum_{k=1}^{x} I^2(x,y)$ is the integrated back scatterer map along the row direction for each pixel (x,y) and $\sigma_0$ is the bandwidth which is fixed to be 0.55. The input to the CNN is the acoustic feature fusion image, $T(I, \lambda_0)$. Thus a 10-channel T(I) is employed, where 10-channels are formed by varying the wavelength parameter, $\lambda_0$, of the log Gabor filter (step 2.a in Fig. 2) [22]. As can be seen from the sub-plots in Fig. 1, the different channels are equivalent to different narrow band acoustic feature representations of the given LUS frame.

**KeyNet regression:** A KeyNet similar to [20] is employed to regress the keypoints in form of $k$ gaussian heatmaps, where $k$

1. T = function (I,IBS,$\sigma_0$, thresh)
2. for $\lambda_0$= 3 to 30
    a. $G(\omega) = exp\left(-\frac{\left(log\left(\frac{|\omega|}{\omega_0}\right)\right)^2}{2(log(\sigma_0))^2}\right)$, where $\omega_0 = 2\pi/\lambda_0$
    b. Calculate monogenic signals, $m_1, m_2$ and $m_3$ of I as in [22]
    c. Calculate even = $|m_1|$ and odd = $\sqrt{m_2^2 + m_3^2}$
    d. Calculate local phase, LP(x)=1- $\tan^{-1}\frac{\sqrt{m_2^2+m_3^2}}{m_1}$
    e. Calculate phase symmetry, FS(x)= $\frac{max(even-odd-thresh)}{m_1^2+m_2^2+m_3^2}$
    f. Calculate feature map, T(I,$\lambda_0$)=LP(x)×FS(x)×(1-IBS)
3. Return T(I,$\lambda_0$)

Fig. 2 Steps in the acoustic feature fusion algorithm

are the number keypoints. As can be seen from Fig. 1, keypoints essentially highlights the essential LUS land marks (Pleura and B lines in the example shown in Fig. 1). It has to be noted that when compared to [20], the proposed frame work reconstructs keypoints for each of the 10 wavelengths.

**Transportation:** As in [20], The keypoints of the source and target images along with the FF-CCN features maps are further used to transport features from source image to target image.

**RefineNet:** A RefineNet [20] is employed to refine the transportation carried out earlier by cleaning up the regions target keypoints and inpainting the region in source keypoints to reconstruct the 10 wavelength channels with higher resolution.

In order to minimize faulty training where various LUS landmarks disappear or emerge between the frame pairs, each frame pair sample drawn from the video data is checked for a similarity threshold with the usage of standard structural similarity index (SSIM). The threshold for grouping as a pair of source and target images was determined to be 0.85 through an ablation study. The network was trained for 60 epochs for 10 keypoints with Adam optimizer with a learning rate of 0.001 (decayed by 0.95 every 6 epochs) with a batch size of

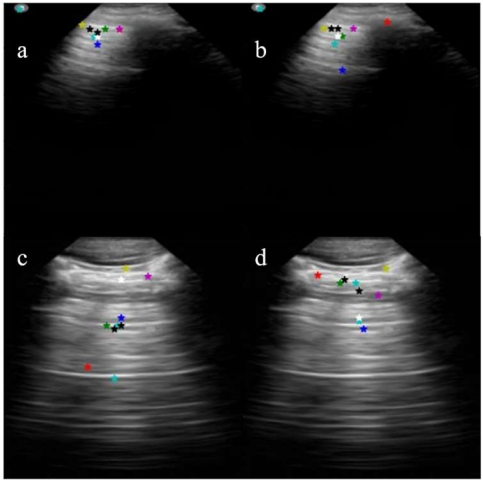

Fig. 3. (a, c) Keypoints generated based on 10 varying (μ, σ) based normalized images (b, d) Keypoints generated based on 10 varying wavelengths based acoustic feature fusion images. Detection performance in (a, c) is similar to (b, d)

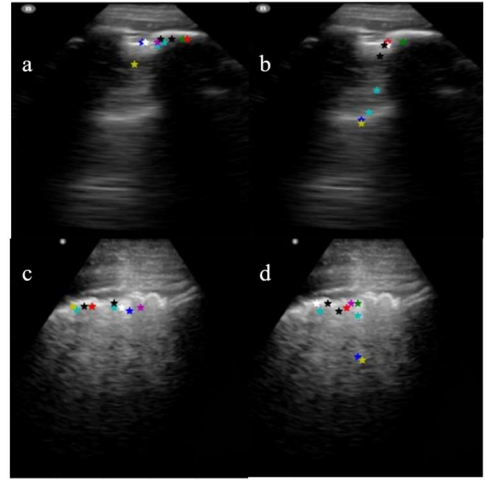

Fig. 4. (a, c) Keypoints generated without considering SSIM between frames (b, d) Keypoints generated by considering only frames with SSIM > 0.85. In (b, d), the A-lines and Blines are identified compared to (a, c)

32. The rest of the hyper parameters were adopted from [20]. During evaluation phase, each LUS sample is forward passed through the KeyNet (as shown in dotted lines as inference mode in Fig. 1) and every $i^{th}$ output channel up sampled from 64x64 to 256x256 in order to plot the ith key point on the input LUS frame.

### III. RESULTS AND ANALYSIS

In this section, the details of the dataset employed for the analysis of the proposed approaches and also the analysis results are presented.

*A. Dataset Description*

In this work, 73 lung ultrasound videos acquired using different ultrasound machines (Butterfly network. GE, Philips and Fujifilm Sonosite with 2MHz center frequency and Lung preset) are employed for training. The videos comprise of healthy lung, lung with B-lines and various stages of lung consolidation during COVID-19. The data acquisition points are anterior, lateral and PLAPS locations on both left and right sides as per the protocol in [2] resulting in a total of six acquisition points.

*B. Ablation Studies*

The ablation studies are conducted in a progressive manner as below:
- To show the significance of acoustic feature maps.
- To show the significance of SSIM.
- To show the significance of the TGA.
- To show the significance of the convolutional block attention module (CBAM).

In the first study, we compare the evaluation results obtained with 10 channels of varying wavelength acoustic feature fusion maps and 10 channel of gray scale images with varying levels of normalization. For the latter, normalization with standard deviation (σ) of 0.5 and mean (μ) varied linearly between 0.3 and 0.7 with respect to the index of the channel are employed to form 10 input channels for the FF-CNN. The results for this ablation study, as shown in Fig. 3, shows that the normalization process filters out the unnecessary features of the LUS images to an extent and results are comparable with that of acoustic feature fusion. In the second ablation, we aim to know the improvement in the results caused with the addition of SSIM refinement to the training process while evaluating the results with acoustic feature fusion. As can be seen from the results in Fig. 4, it is evident that the SSIM refinement proved to have made the training process more stable in terms of tracking the prominent LUS landmarks like pleura and A lines. In the next study, we incorporated CBAM module [23] in order to filter out the irrelevant convolutional channels and spatial positions for better key point formation and trained the transporter with SSIM refinement and evaluated the results with the usage of acoustic feature fusion. As evident from the results shown in Fig. 5, the results with the usage of CBAM are better in terms of tracking the more prominent features with keypoints. However, in images with bright B patches, the keypoints were more clustered and appear to track the bright B patches in an unsteady manner. Hence the CBAM's effects are meant to be studied thoroughly as a future work. In the final study, we incorporate both TGA and SSIM refinement to bias the transporter network toward pleural detection by utilizing the spatial uniformity expected to be observed in a clinical LUS. The results, as shown in Fig. 6, are far more stable in terms of pleural and A line tracking in cases with bright B patches. In order to quantify the performance of the proposed unsupervised key point learning approach, the accuracy of detecting most common landmark like pleura has been evaluated. The results show that the proposed approach has been successful in correctly detecting pleura in 950 frames out of 1081 frames which shows an accuracy of 91.8%.

### IV. CONCLUSIONS

Automated workflows for lung ultrasound imaging have been gaining popularity as a continuous monitoring tool for lung infections. However, most of the works in literature try to detect or segment the pre-determined key landmarks employing fully or semi-supervised learning approaches. In this work, we try to leverage the learning power of neural networks to track the key landmarks in an unsupervised manner. Towards this, a novel unsupervised learning framework is proposed, which leverages the transporter architecture in order to learn the key point representations for tracking prominent features based on their periodic motion and relatively stable appearance in the video sequences. On applying to the task of unsupervised pleura detection, the proposed approach shows a very promising performance of 91.8% accurate landmark detection over 1081 LUS video frames randomly selected from 73 videos.

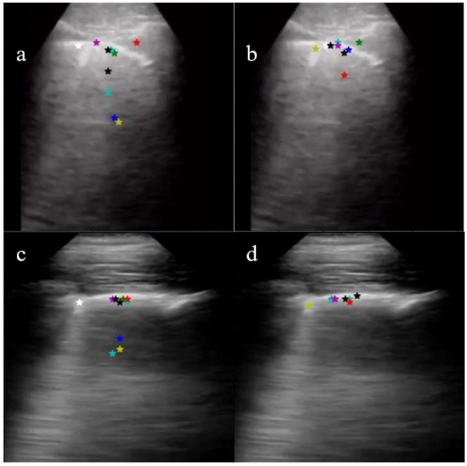 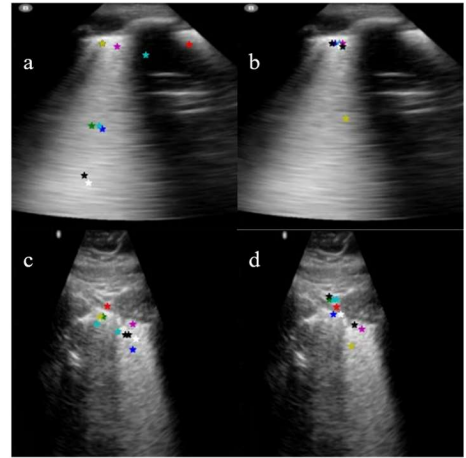

Fig. 5. (a, c) Keypoints generated without CBAM (b, d) Keypoints generated by considering CBAM

Fig. 6. (a, c) Keypoints generated without TGA and SSIM (b, d) Keypoints generated by considering TGA and SSIM